\pdfoutput=1

\documentclass[10pt]{article} 
\usepackage{ragged2e} 
\usepackage[utf8]{inputenc} 

\usepackage{geometry} 
\usepackage{graphicx} 
\usepackage{amssymb} 
\usepackage{amsmath}
\usepackage{hyperref} 
\usepackage{mathtools}
\usepackage{upgreek} 
\usepackage[parfill]{parskip} 
\usepackage{float} 

\usepackage{booktabs} 
\usepackage{array} 
\usepackage{paralist} 
\usepackage{verbatim} 
\usepackage{subfig} 
\usepackage{tablefootnote} 

\usepackage{fancyhdr} 
\usepackage{sectsty}
\allsectionsfont{\sffamily\mdseries\upshape} 

\usepackage[nottoc,notlof,notlot]{tocbibind} 
\usepackage[titles,subfigure]{tocloft} 

\title{JWST MIRI Imaging Data Post-Processing Preliminary Study with Fourier Transformation to uncover potentially celestial-origin signals}
\author{Güray Hatipoğlu}

\begin{document}
\maketitle

\centering{\section*{Abstract}}
\justifying

This manuscript reports a part of a dedicated study aiming to disentangle sources of signals from James Webb Space Telescope (JWST) Mid-Infrared Instrument (MIRI) imaging mode. An instrumental introduction and characteristics section is present regarding MIRI. Later, a Fast Fourier Transformation-based filtering approach and its results will be discussed.

\setlength{\columnsep}{3em}
\setlength{\columnseprule}{1pt}

\section*{JWST Mid-Infrared (MIRI) Instrument}

The wavelength range James Webb Space Telescope (JWST) Mid-Infrared Instrument (MIRI) works is from 4.9 to 27.9 $\mu$m. MIRI has four different modes to utilize the information coming in this range: imaging (nine photometric bands), low-resolution slit and slitless spectroscopy, medium-resolution spectroscopy with four different integral field units (IFU), and coronagraphy with four filters \cite{stsci2016}. The following subsection will provide more details on the imaging mode.

\subsection*{\textit{Imaging Mode Features}}

The MIRI Imager covers 5.6 to 22.5 micrometer wavelength range with a clear field of view in 74'' x 113'' dimensions. All filters are diffraction limited, wavelengths shorter than 15 micrometers have astronomical background limited sensitivity, while filters smaller than 15 micrometers wavelength have a primary mirror and sun shield-related telescope background limitation. The following table summarizes the filter properties in MIRI. 

\begin{center}
\centering\captionof{table}{MIRI filter properties \cite{stsci2016}}
\begin{tabular}{c|c|c|c|c}
\hline

Filter & $\lambda_{0}$ ($\mu$m) & $\Delta\lambda_{0}$ ($\mu$m)  &  FWHM (arcsec) $^{1}$ & Comment\\
\hline
 F560W& 5.6& 1.2& 0.207& broadband imaging, undersampled\\
 F770W& 7.7 & 2.2 & 0.269& PAH,  broadband, Nyquist sampled\\
 F1000W& 10.0& 2.0& 0.328& silicate,  broadband, oversampled\\
 F1130W& 11.3& 0.7& 0.375&  PAH,  narrower broadband, oversampled \\
 F1280W& 12.8& 2.4& 0.420&  broadband, oversampled\\
 F1500W& 15.0& 3.0& 0.488&  broadband, oversampled\\
 F1800W& 18.0& 3.0& 0.591&  silicate,  broadband, oversampled\\
 F2100W& 21.0& 5.0& 0.674&  broadband, oversampled\\
 F2550W& 25.5& 4.0& 0.803&  broadband, oversampled
\label{Table. 2} 
\end{tabular}
\begin{tablefootnote}
\item Full-Width at Half Maximum from Point Spread Function (PSF)
\end{tablefootnote}

\end{center}

\justifying

Parameters to set for MIRI imaging observation are filters, dither pattern, subarray choice, detector read-out modes, and exposure time (with the number of frames and integration). Various subarray modes exist to handle very bright sources or backgrounds (Table 2).

\begin{center}
\centering\captionof{table}{MIRI subarray characteristics \cite{stsci_2016b}}
\begin{tabular}{c|c|c|c}
\hline

Subarray & Size (pixels) & Usable size (arcsec) & Frame time \\
\hline
FULL & 1024 x 1032& 74'' x 113''& 2.775 s\\
BRIGHTSKY & 512 x 512& 56.3'' x 56.3''& 0.865 s\\
SUB256 & 256 x 256& 28.2'' x 28.2''& 0.300 s\\
SUB128 & 128 x 136& 14.1'' x 14.1''& 0.119 s\\
SUB64 & 64 x 72& 7'' x 7''& 0.085 s
\label{Table. 2} 
\end{tabular}

\end{center}

Dither options allow oversampling the PSF, removing bad pixels, and optimizing the self-calibration process.

\section*{Aim}

JWST MIRI images seem to have several spatially periodic patches, patterns, and shapes that might be non-celestial in origin and may mask actual celestial objects. Fourier Transformation is an established method to filter out or extract periodic signals from time-series or 2-dimensional image data. This paper describes a preliminary approach to post-process calibration level 3 JWST MIRI images.

\section*{Methodology}

The study mainly utilized two different methods to process the data, which was in detail in the "Method" subsection. Python 3.x environment and Astropy (to open FITS images), NumPy (to conduct FFT), and Matplotlib (to plot the results) packages are necessary and sufficient to make a similar analysis. The convenience functions for this case can be found in \url{https://github.com/torna4o/fft_for_jwst}.

\subsection*{Background}

Fourier series is a representation of any type of periodic signal with an infinite number of sinusoidal components\cite{bevelacqua}:

\[
     g(t) = \alpha_{0}+  \sum_{m=1}^{\infty}\alpha_{m}cos(\frac{2\pi mt}{T})+
    \sum_{n=1}^{\infty}\beta_{n}sin(\frac{2\pi nt}{T})
\]

Such Fourier series expansion can represent \textit{continuous} and \textit{smooth} data. The complex representation is as follows:

\[
     g(t) = \sum_{n=-\infty}^{\infty}c_{n}e^{t\frac{2\pi nt}{T}}
\]

The optimum values for constat $c_{n}$ are:

\[
c_{0} = \frac{1}{2}    
c_{n} = \frac{1}{i\pi n}, n=\pm1,\pm3,\pm5,...
c_{n} = 0, n=\pm2,\pm4,\pm6,...
\]

The Fourier transformation method expands this approach to non-periodic functions.

\[
F[g(t)] = G(f) = \int_{-\infty}^{\infty} {g(t)e^{2-\pi ift}dt}
\] 

One property of this Fourier Transformation is its linearity; that is, Fourier Transformation of the sum of two arbitrary functions will yield the same result as the summation of individual Fourier Transform of those functions.

This approach directly works with the data using pre-defined filters. Another way is to find the specific frequency of the redundant part of the data programmatically and try to filter it out. This study will use Welch's method \cite{1161901} in a one-dimensional reshaped matrix data and filter the original data in 2-dimensions one by one.

\[
	X_{1}(j) = X(j) 		j=0,1,2,...,L-1,
\]
\[
	X_{2}(j) = X(j+D)  		j=0,1,2,...,L-1,
\]
\[
	X_{K}(j) = X(j + (K-1)D)  j=0,1,2,...,L-1
\]

These will be multiplied by \textit{W} data window and their finite Fourier Transforms will be taken.

\[
A_{k}(n) = \frac{1}{L}\sum_{j=0}^{L-1} {X_{k}(j)W(j)e^{-2kijn/L}}
\]

There are now K modified periodograms:

\[
I_{K}(f_{n})=\frac{L}{U}|A_{k}(n)|^{2}; 	k=1,2,....,K
\]
\[
f_{n} = \frac{n}{L}, n= 0,...,L/2
\]
\[
U=\frac{1}{L}\sum_{j=0}^{L-1} {W^{2}(j)}
\]

The average of K periodograms above gives the spectral estimate.

\subsection*{Data}

The data was from the Mikulski Archive for Space Telescopes Portal. It is publicly available data from ERS Program Category, Stellar Physics science category with the program title of \textit{Radiative Feedback from Massive Stars as Traced by Multiband Imaging and Spectroscopic Mosaics}. The principal investigator is Olivier N. Berne and the Program number is 01288\cite{stsci}. Target right ascension in degree decimals is 81.8308, and the target declination is -5.5345. The readout pattern is FASTR1 and the primary dither pattern is "4-point". 

\begin{figure}[H]
\includegraphics[width=\columnwidth]{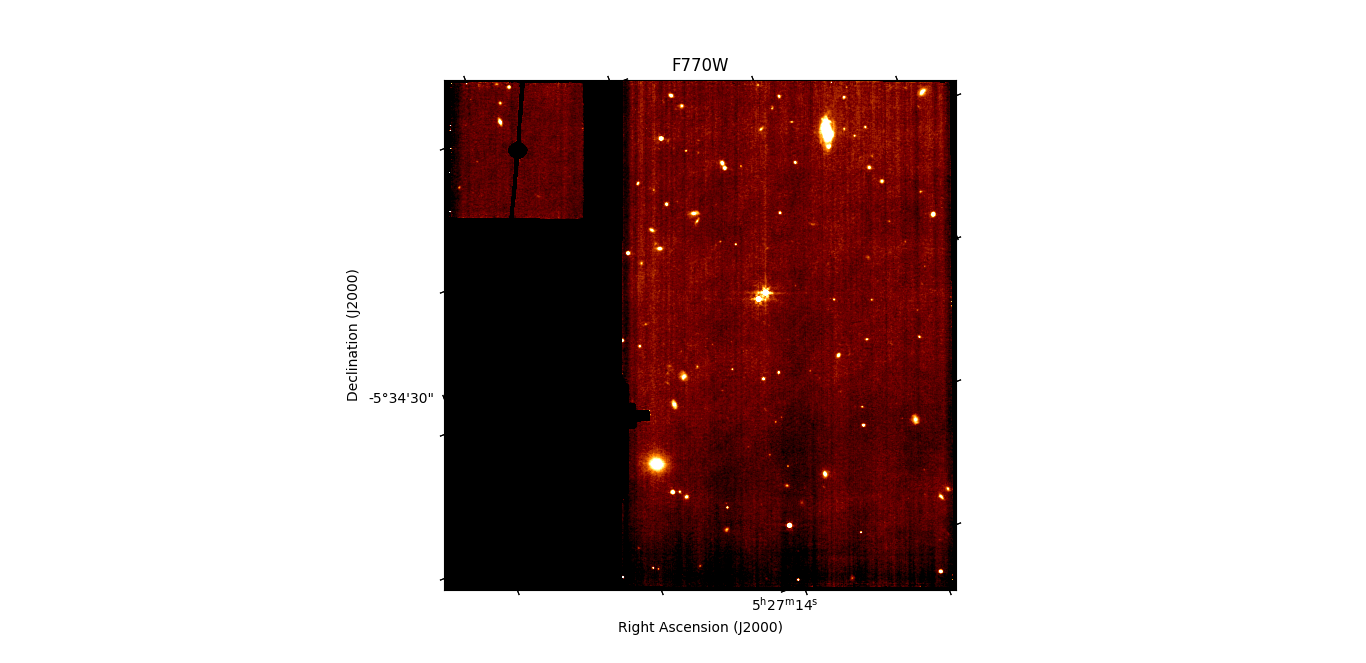}
\includegraphics[width=\columnwidth]{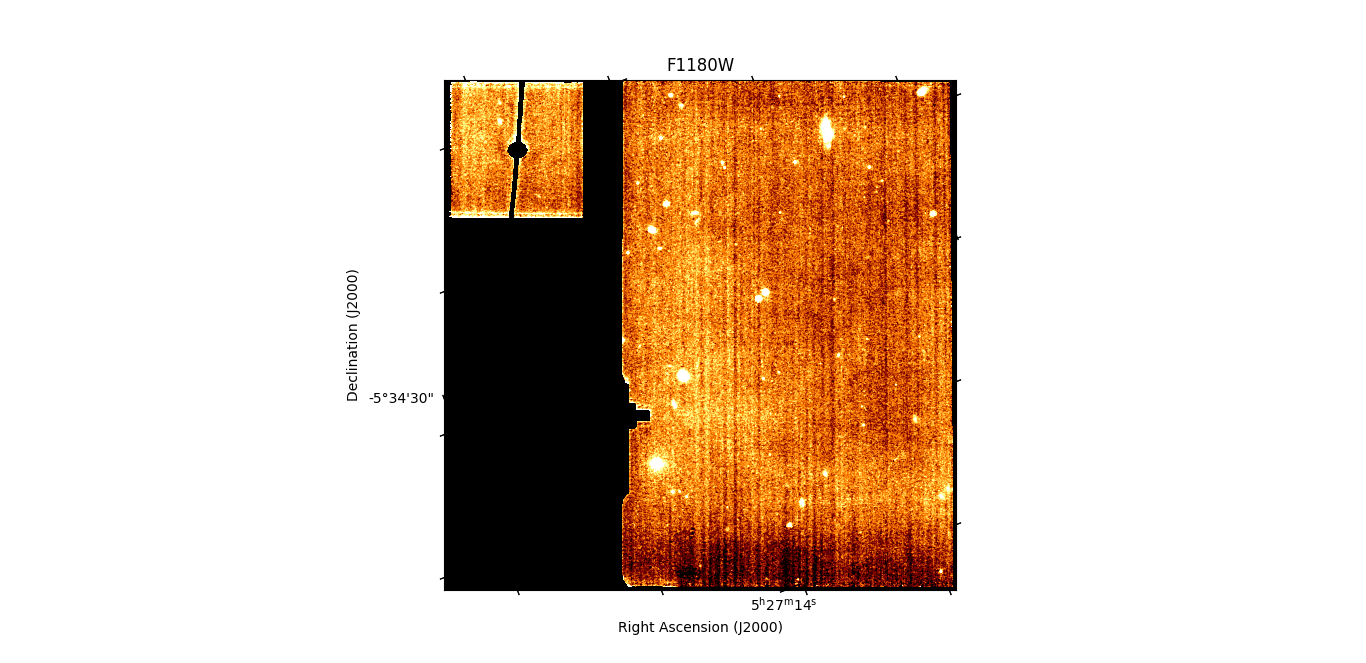}
\includegraphics[width=\columnwidth]{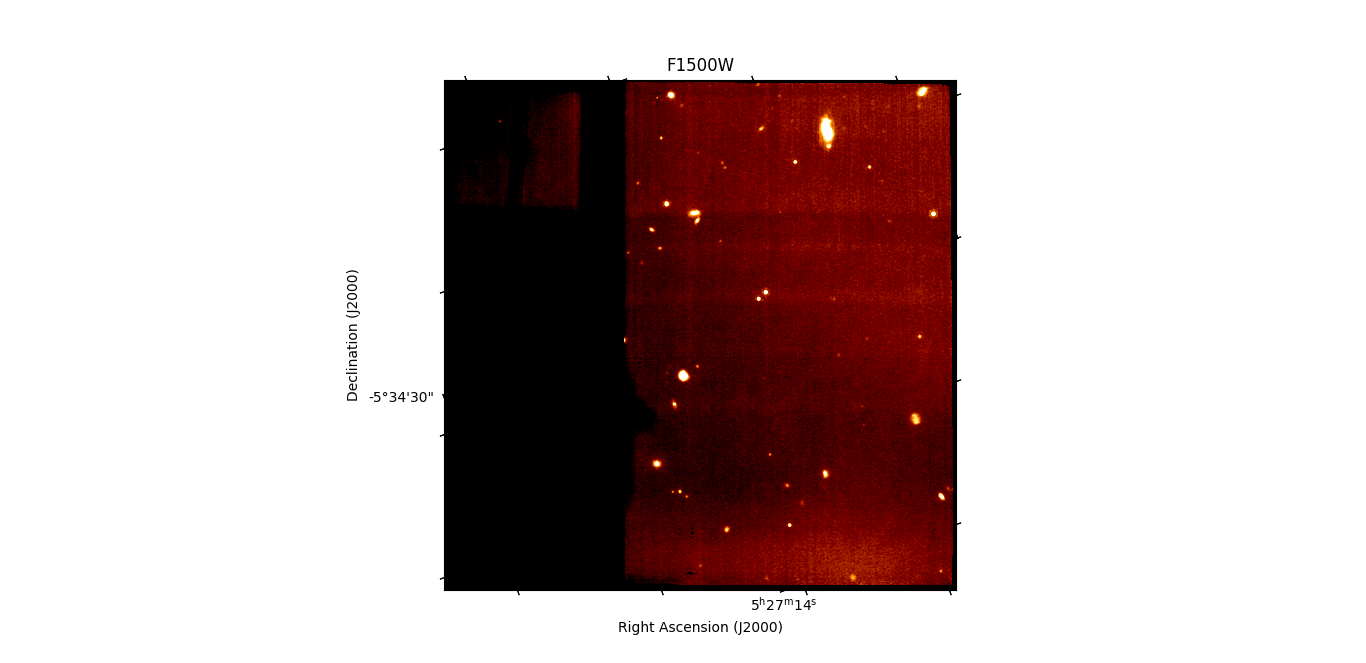}
\caption{JWST-MIRI imaging data in F770W, F1130W, and F1500W filters}
\end{figure}

\subsection*{Method}

There are non-celestial-looking patterns in imaging data (Figure 2). One established way to filter out such data is Fourier Transformation and low-pass or high-pass filter. A part of the image data without any spatial discontinuity was cut out (in 560x780 pixel dimensions). 

\begin{figure}[H]
\includegraphics[width=\columnwidth]{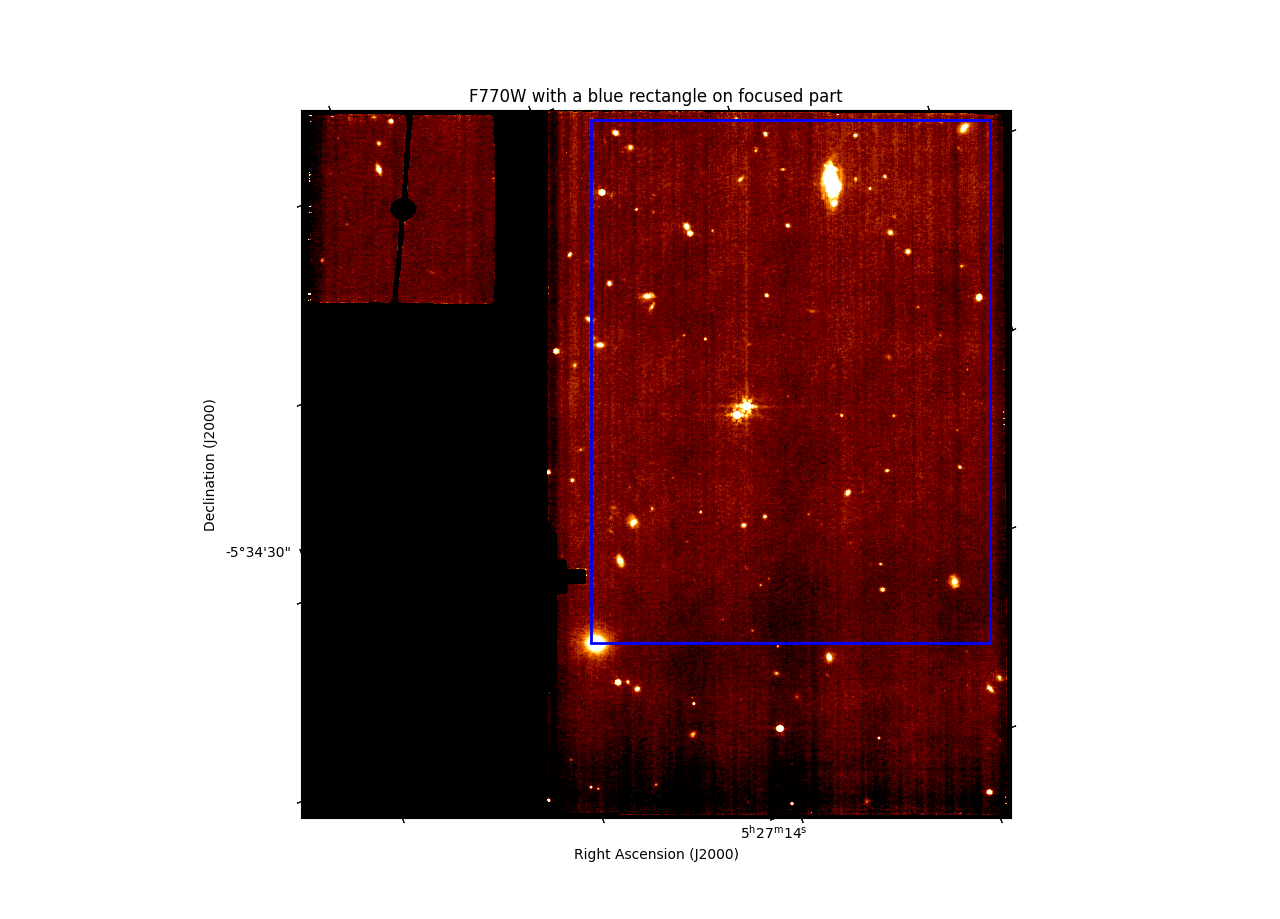}
\caption{JWST-MIRI imaging data of the region of interest in all filters}
\end{figure}

Then, the following Fourier transformation approaches were applied to this image.

\subsubsection*{Spherical Filter-Frequency-based Removal}

\begin{figure}[H]
\includegraphics[width=\columnwidth]{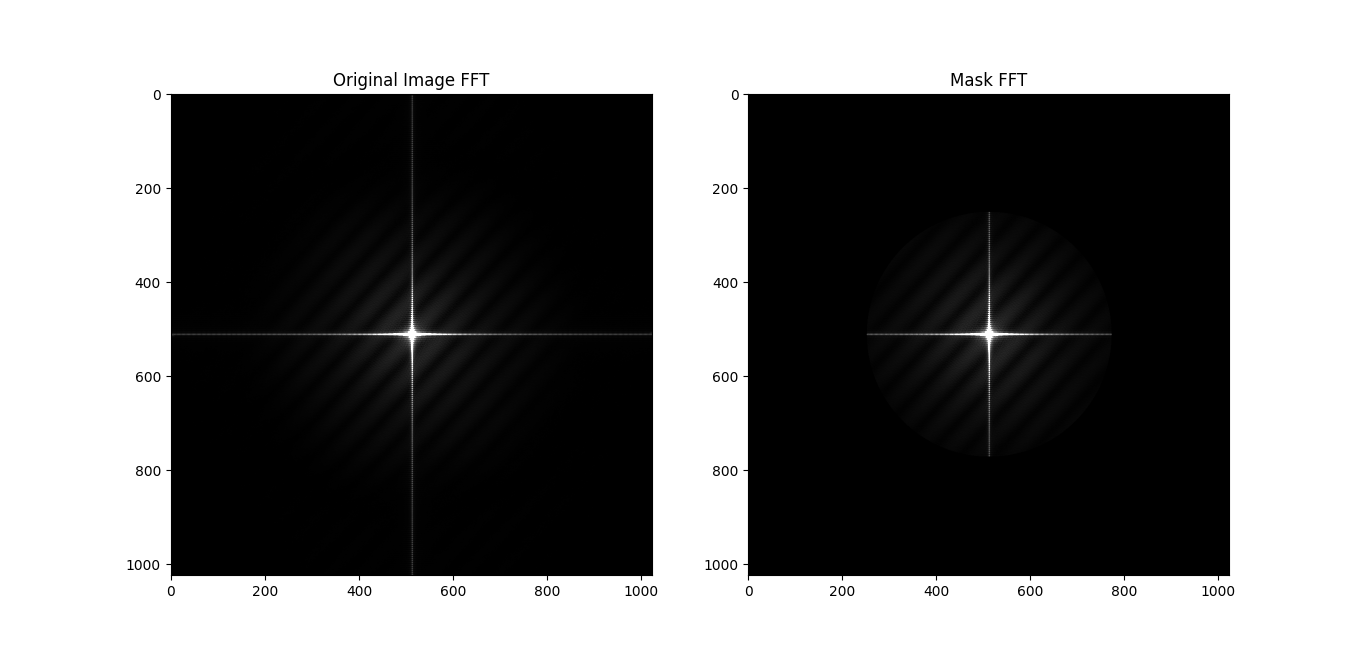}
\caption{Circular frequency mask with an adjustable radius, in this case, 260-pixel radius}
\end{figure}

In this way, the radius of this circle is the only parameter to be chosen. There is also an inverted mask where the data remained after filtering this circular patch is available.
\subsubsection*{Cross Shape Frequency-based Removal}

\begin{figure}[H]
\includegraphics[width=\columnwidth]{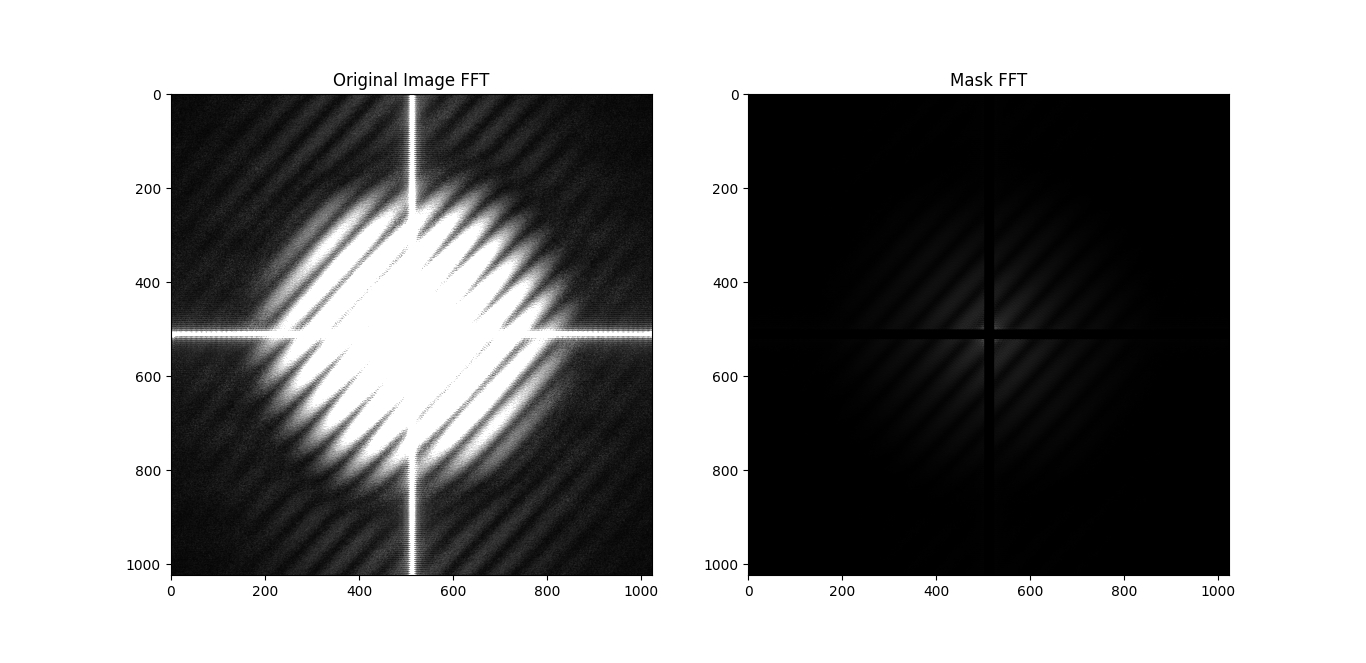}
\caption{Cross shape frequency mask with an adjustable horizontal and vertical line thickness}
\end{figure}

\subsubsection*{Pattern Recognition and Specific Signal Removal}

This approach do not have a predefined filter. Instead, it constructs a filter based on the data. As explained, the code utilizes Welch's method and a Fourier transform-based operation.

\subsection*{Results}

\subsubsection*{Circular Filter-Frequency-based Removal}

\begin{figure}[H]
\includegraphics[width=\columnwidth]{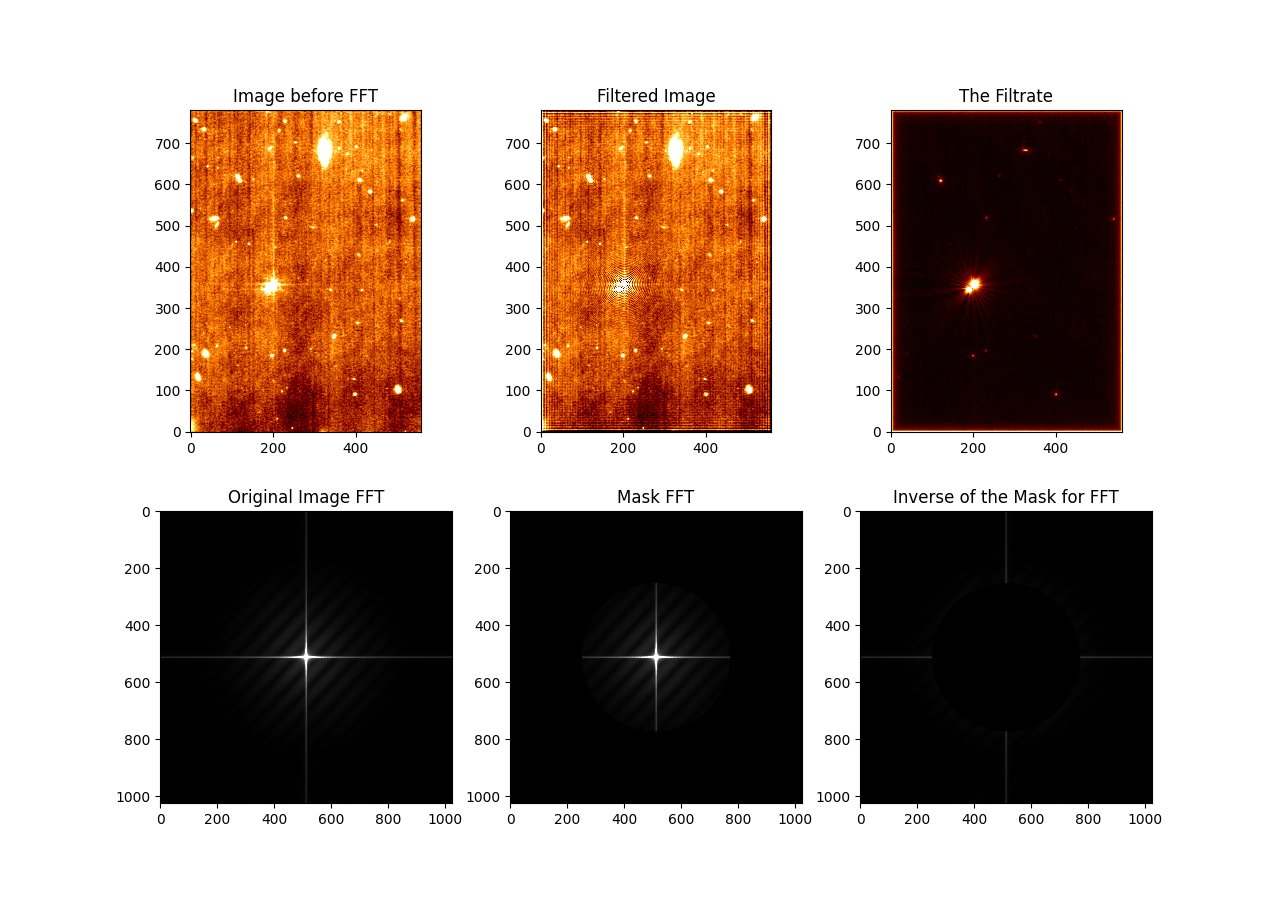}
\includegraphics[width=\columnwidth]{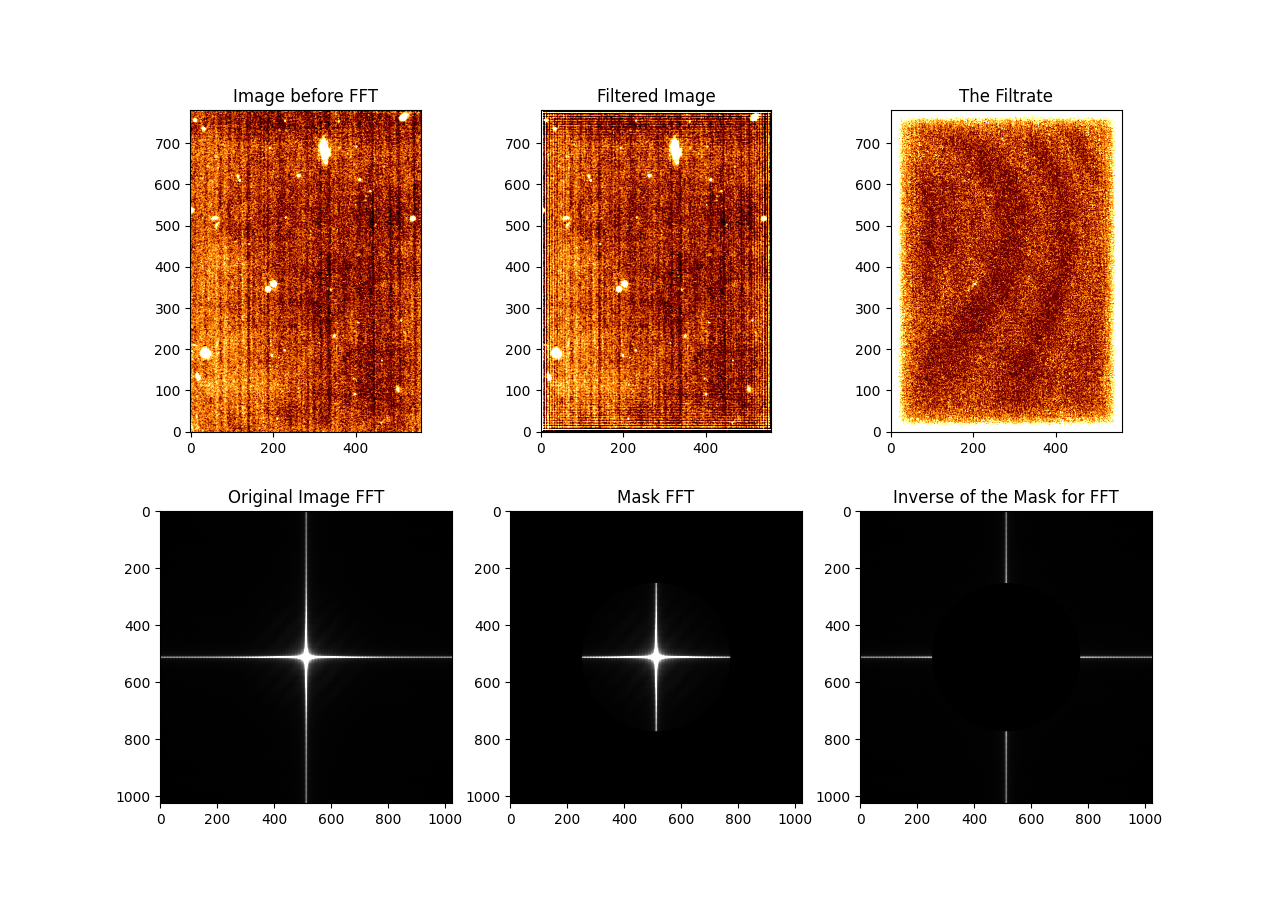}
\caption{Spectral-domain filtering with a circular shape in frequency domain in all filters, from top to bottom, F770W and F1130W}
\end{figure}

\begin{figure}[H]
\includegraphics[width=\columnwidth]{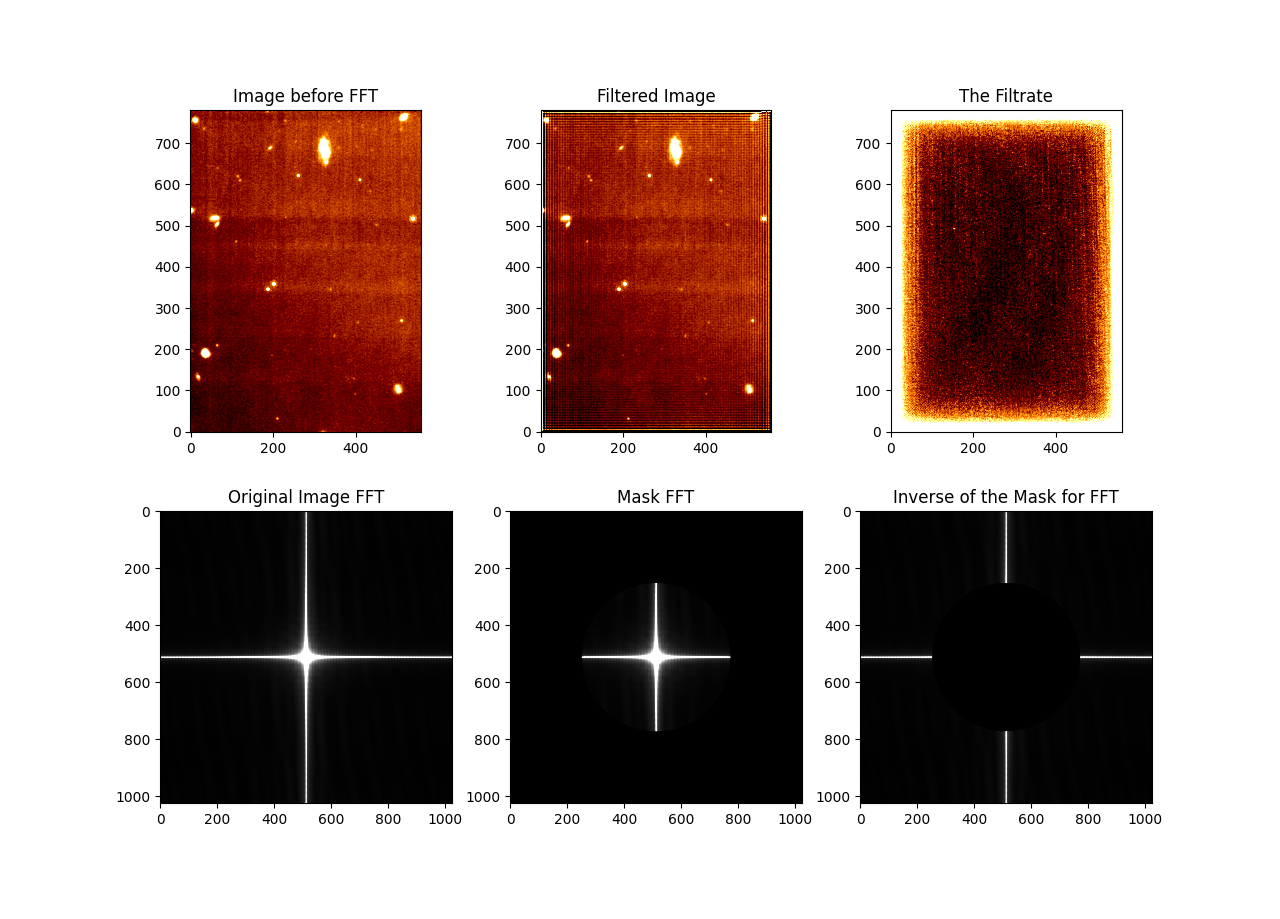}
\caption{Spectral-domain filtering with a circular shape with a threshold of 260 pixels in the frequency domain in F1500W filter}
\end{figure}

\subsubsection*{Cross-shape Removal}
\begin{figure}[H]
\includegraphics[width=\columnwidth]{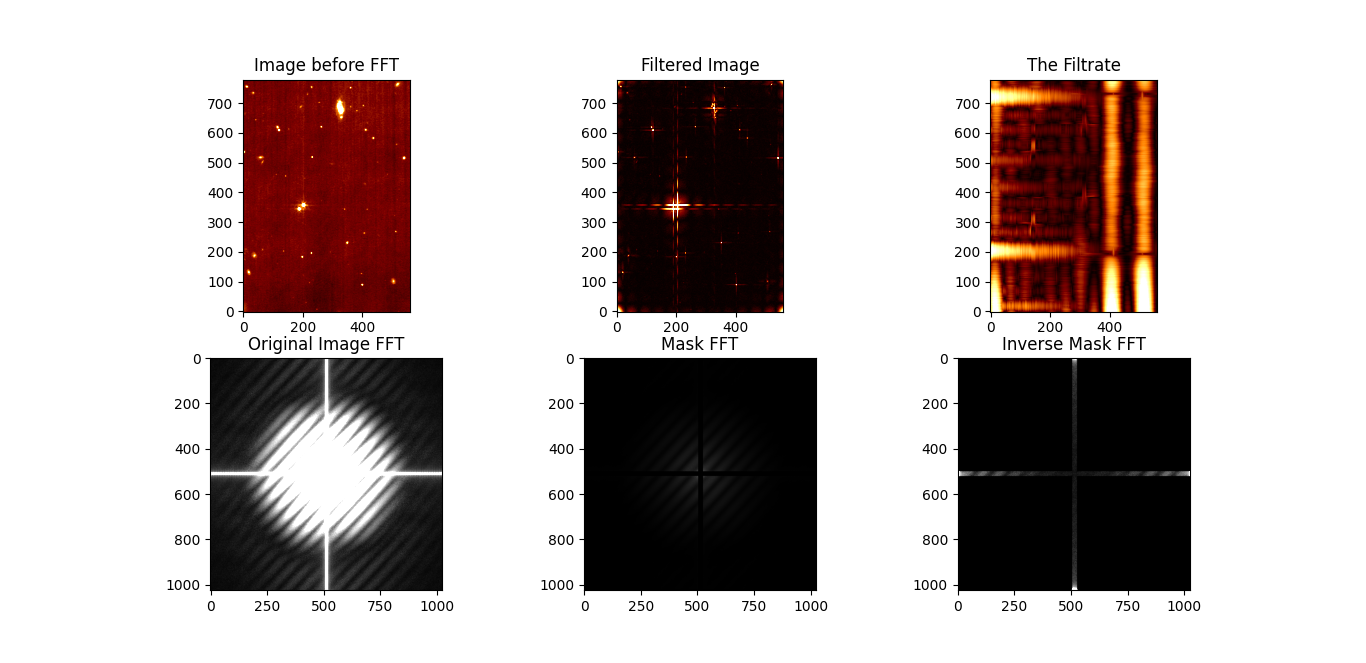}
\includegraphics[width=\columnwidth]{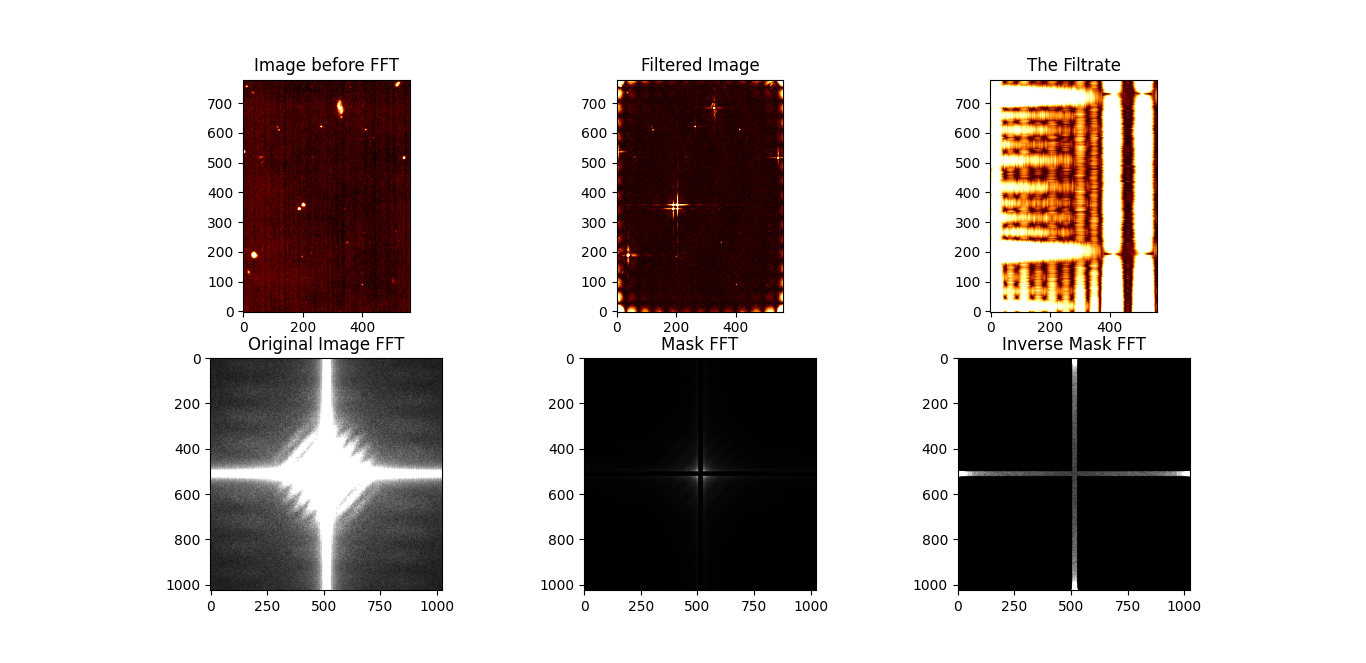}
\includegraphics[width=\columnwidth]{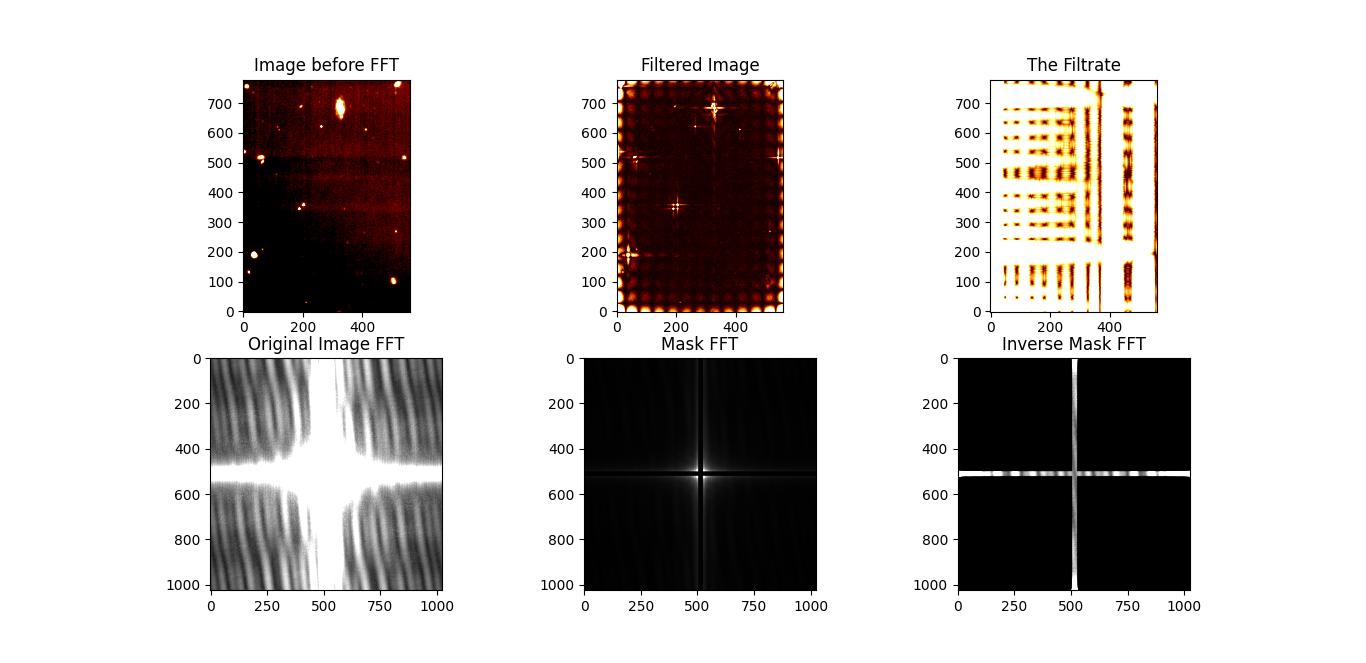}
\caption{Spectral-domain filtering  in all filters with a cross shape, from top to bottom, F770W, F1130W, and F1500W}
\end{figure}

The following figures are from a focused region after cross-shape frequency domain filtering. 

\begin{figure}[H]
\includegraphics[width=\columnwidth]{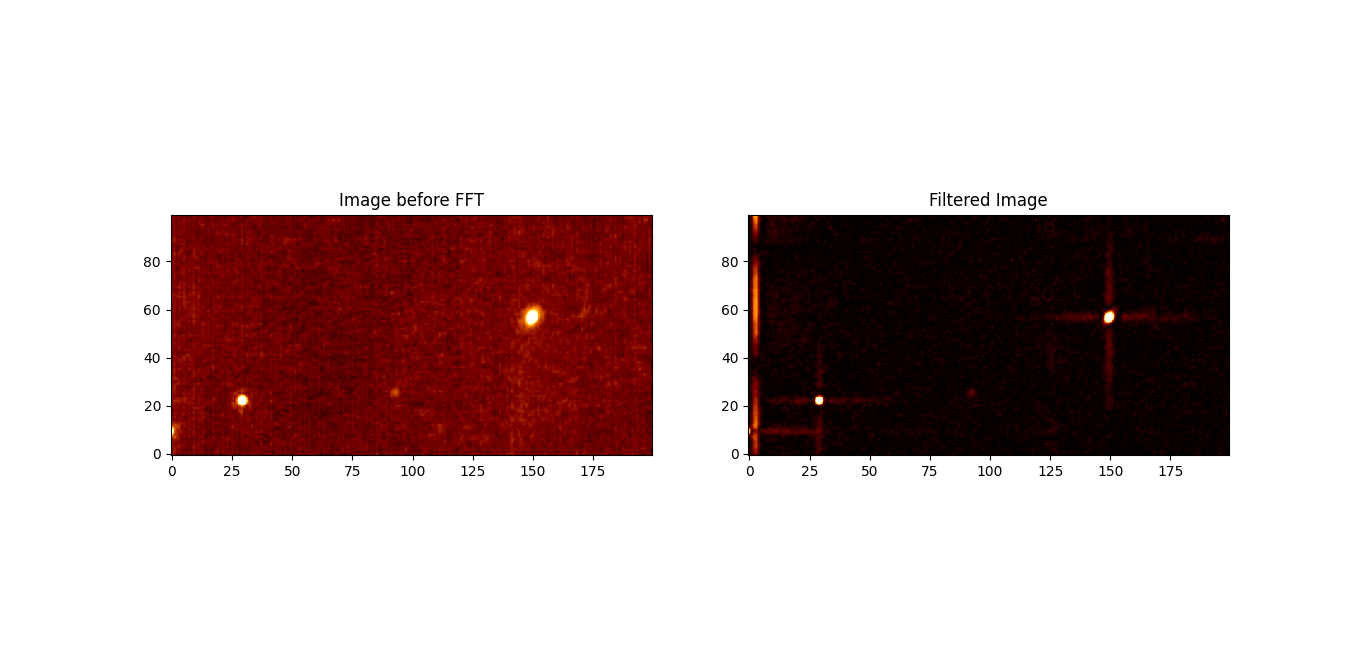}
\includegraphics[width=\columnwidth]{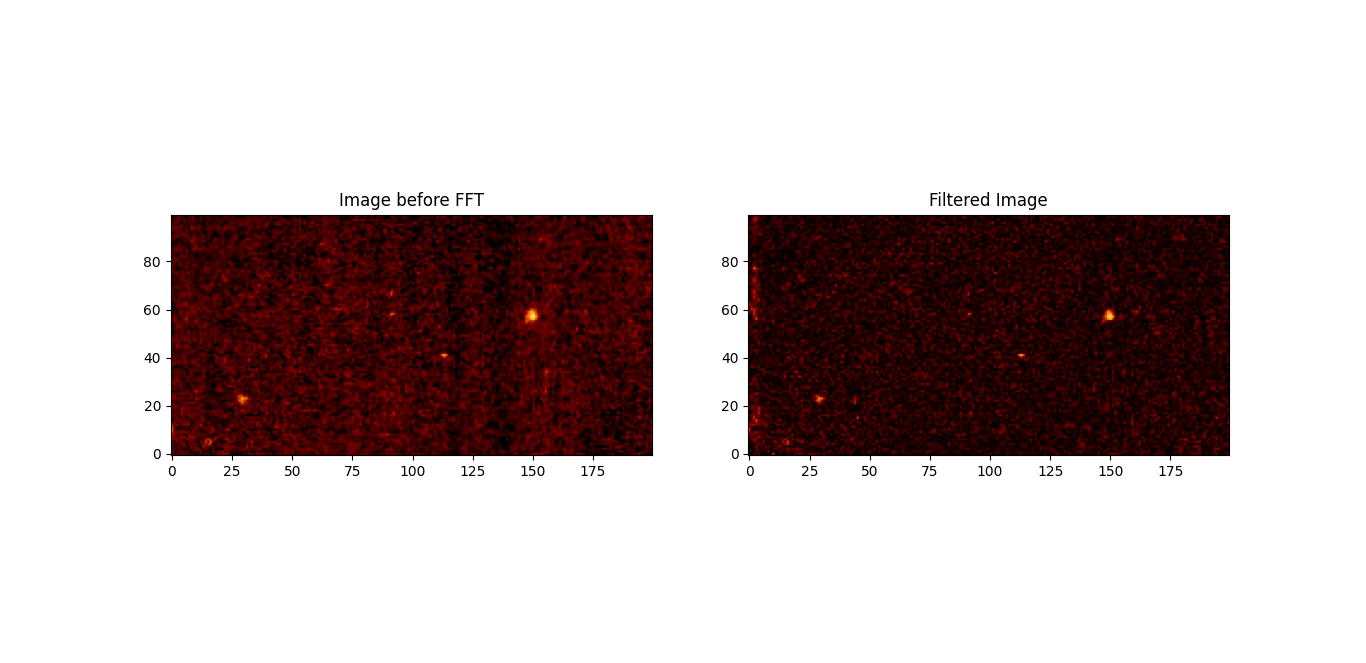}
\caption{A further-focused part of a region to compare the result of a cross-shape masking in spectral dimension for all filters, from top to bottom, F770W and F1130W}
\end{figure}

\begin{figure}[H]
\includegraphics[width=\columnwidth]{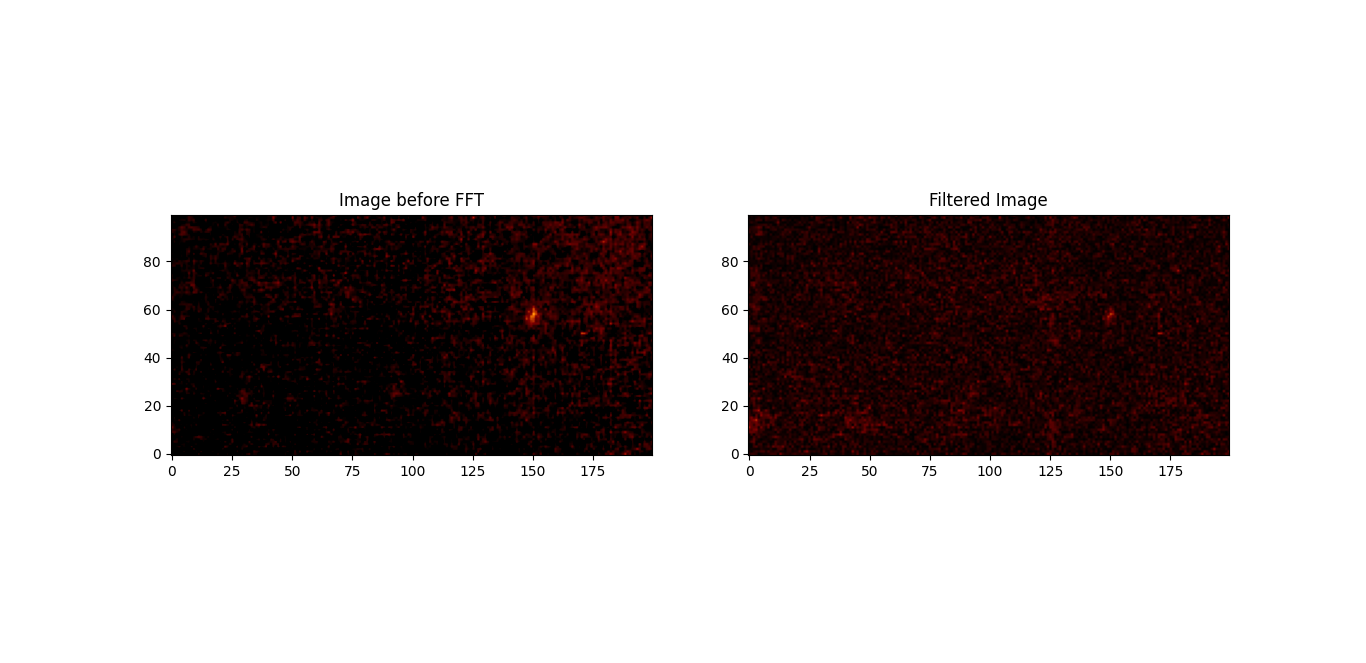}
\caption{A further-focused part of a region to compare the result of a cross-shape masking in spectral dimension for F1500W filter}
\end{figure}

In all cases, the approach removed the vertical strips present in the data prior to the Fourier Transformation. However, only in the F1130W filter, there are visible impacts and potential objects uncovered.

\subsubsection*{Pattern Recognition with Welch Method and Specific Signal Removal}

\begin{figure}[H]
\includegraphics[width=\columnwidth]{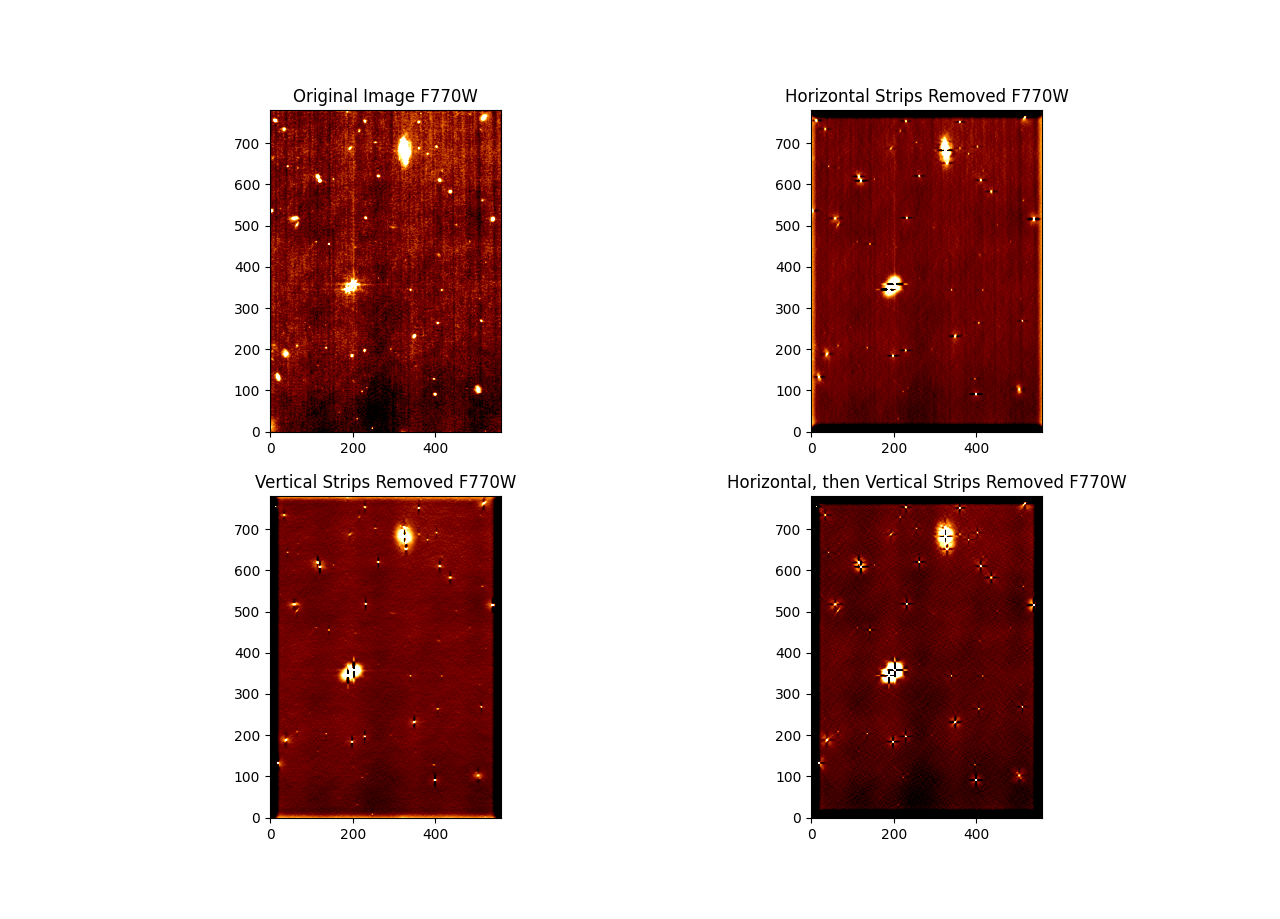}
\includegraphics[width=\columnwidth]{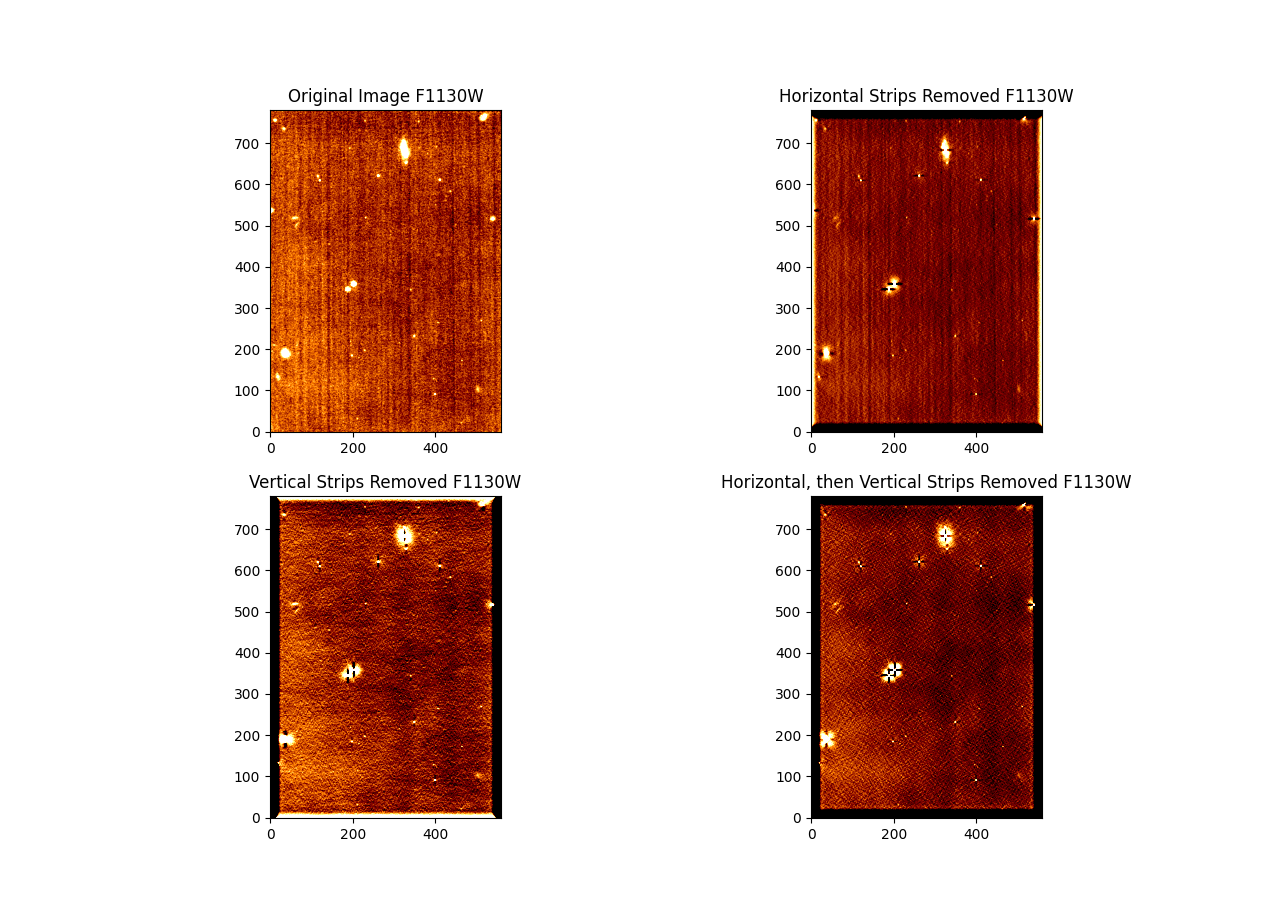}
\caption{A further-focused part of a region to compare the result of the specific signal removal in spectral dimension for all filters, from top to bottom, F770W and F1130W}
\end{figure}

\begin{figure}[H]
\includegraphics[width=\columnwidth]{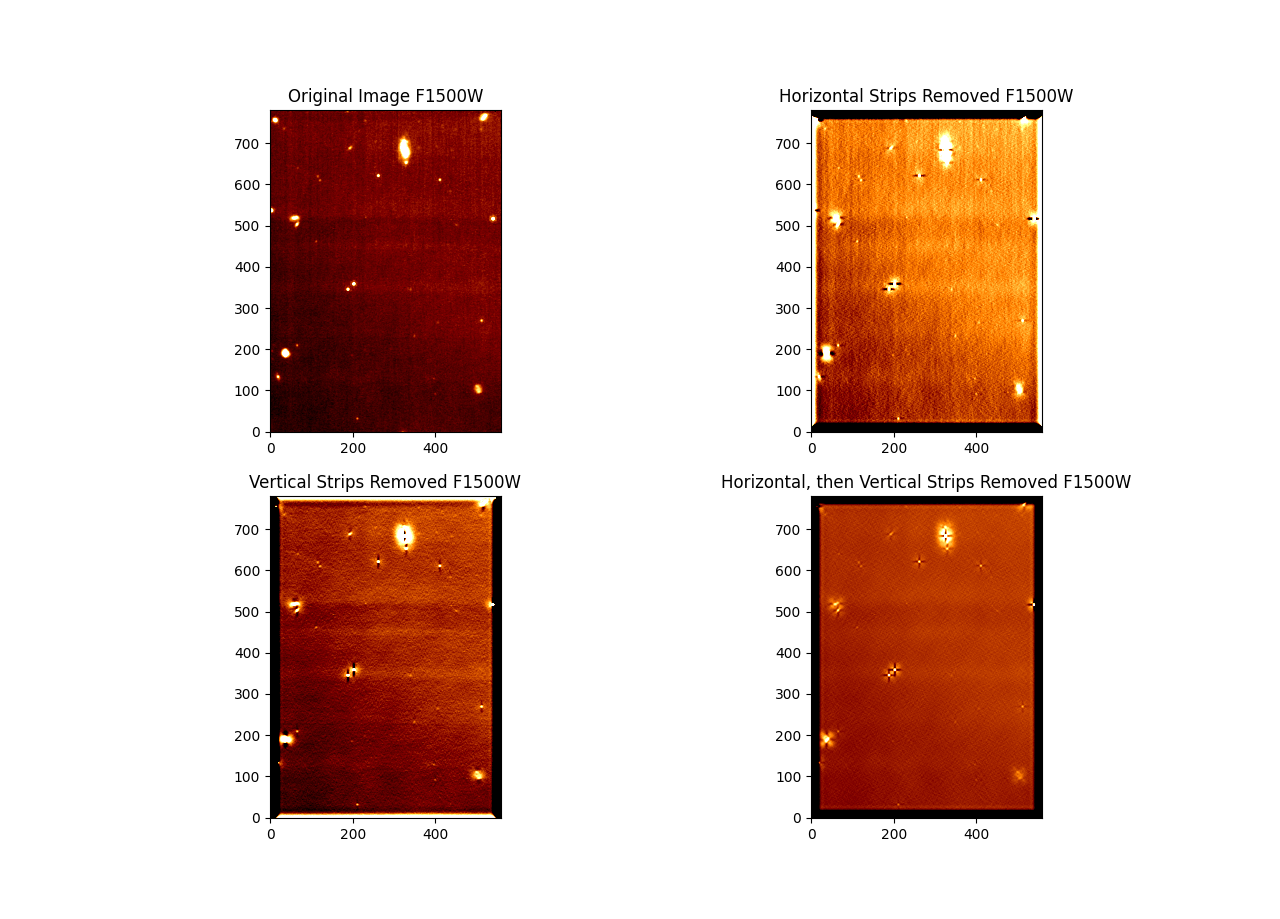}
\caption{A further-focused part of a region to compare the result of the specific signal removal in spectral dimension for F1500W filter}
\end{figure}

\subsection*{Discussion}

The aim was to remove the strips present in the data to uncover or amplify the signal from genuine celestial objects. The preliminary trials gave somewhat promising results to more precisely discover the signal responsible for the vertical (all filters) and horizontal (in F1500W, especially), strips, and then, extract it from the images.

A circular filter with 260 pixel-radius threshold reduced the vertical strips in the image greatly (Figure 4), and completely eliminated thick horizontal strips (Figure 5). Nevertheless, a new wavy pattern across the image in a vertical direction was introduced in the inverted mask images (Figures 4 and 5). Only for F1500W, there were several points with suspected signals (Figure 5, right top "The Filtrate" image location: x=$\sim$175, y=500).

Since the data has a distinct cross-shape component in the frequency domain, a cross-shape mask was also applied to see its results. In front of a wavy pattern encountered in the circular-shape mask case (Figures 4 and 5), a seemingly easy-to-reproduce grid pattern was introduced in the filtered image after the operation, with higher magnitude in corners and sinusoidally-decreasing magnitude towards the center. However, the operation did remove the initially present strip patterns, and especially for the F1130W filter, (Figure 7, bottom), several more points in the image appeared that might be of interest (location: x=$\sim$ 45, y=20).

Pattern recognition-related approach, indeed, removed most or all horizontal and vertical strips in the map. However, this was at the expense of star signal magnitudes and the number of bright pixel neighbors, which will most probably impact source detection algorithm results. There are also bad margins in the direction in which the method was applied, and in the vertical and horizontal direction radii of major brightness sources. Even if not like in previous methods, this approach also resulted in some signal loss while subtracting the pattern (Figures 10 and 11).

\section*{Concluding Remarks}

Several caveats are present to consider while applying FFT and interpreting the results. The first one is the impact of edges in the image to analyze. This was the reason for not ingesting the entire image, but cropping a specific part without any spatial discontinuity. However, the image itself is not fully composed of specific patterns, and rotated patterns have rather complex Fourier Transformation results\cite{brayer}. Thus, examining solely the frequency domain image is challenging to filter out non-celestial signals. A method to examine the expected frequency domain magnitude and phase of a specific signal in the original image can be the following step to reap the benefits of FFT. Another way of solving this issue is trying other methods, such as Discrete Cosine Transformation (DCT) \cite{1672377}, which is better at dealing with edges in discrete signals, intuitively for pixels of an astronomical image.

Another thing to consider is that the FFT does not completely restore all the signals of the original image.  Whether this will result in considerable information loss should be checked.

The final thing to consider is the remaining strips, lines, and patterns, probably in inclined directions. At least for these methods, repeated application severely \textit{dims} the image. This is another issue to be solved in future studies.

\bibliography{miribib} 

\end{document}